\documentclass{llncs}

\usepackage{graphicx}
\usepackage{authblk}
\usepackage{amssymb}
\usepackage{float}
\usepackage{algorithm}
\usepackage{algpseudocode}

\begin{document}

\title{Systematization of a 256-bit lightweight block cipher Marvin }

\author{Sukanya Saha\inst{0000-0003-3995-4752}, Krishnendu Rarhi\inst{0000-0002-5794-215\texttt{x}}, Abhishek Bhattacharya\inst{1}}

\institute{School of Information Technology, unit of Institute of Engineering and Management,Kolkata\\
\email{sukanyash441@gmail.com},\\ \email{rarhik@gmail.com},\\ \email{abhishek.bhattacharya@iemcal.com}}
\maketitle

\begin{abstract}

In a world heavily loaded by information, there is a great need for keeping specific information secure from adversaries.
The rapid growth in the research field of lightweight cryptography can be seen from the list of the number of lightweight stream as well as block ciphers that has been proposed 
in the recent years. This paper focuses only on the subject of lightweight block ciphers. In this paper, we have proposed a new 256 bit lightweight block cipher named as Marvin, that belongs to the family 
of Extended LS designs.

\end{abstract}

\section{Introduction}\label{sec:Introduction}

The emergence of development of various families of ciphers,like  the SPN,that is Substitution Permutation Network,Feistel Network,Generalized Feistel Network\\ contribute to the further development of many lightweight block ciphers.To name some of them, there are, lightweight block ciphers like ,Midori, Zorro,mCrypton,\\CLEFIA,MANTIS,SIMON,SPECK and many more.

These
 lightweight block ciphers are  in todays world,used for various ASIC and RFID tagging systems.
 The above mentioned lightweight block ciphers follow some specific structures,block sizes, number of rounds,etc.
Moreover, all these ciphers are public key,or symmetric block cipher which are used for lightweight cryptography.There are numerous attacks that has been applied to many block ciphers in the past which gradually contributed to the development of more and more complex and secure block ciphers.After the second world war, the design of various ciphers in cryptography has became more mathematically involved. And today, the various fields that are applied in the design of block ciphers are mathematical Probability theory, knowledge of Group theory, Rings ,Field, Discrete Abstract Algebra and Matrix manipulation methods.The field of Coding theory also have  enormous use in making of various components of an algorithm.The lightweight cryptography focuses on the  design simplicity of the ciphers which can provide the opportunity of these ciphers for being used in Internet of Things devices.

\subsection{What is Cryptography?}

Cryptography, is the science of encrypting secret information so that the information can be kept hidden from some intended adversaries.It is that field of study in computer science, 
which focuses in the making of algorithms and structures that can provide protection to information.It also analyses these algorithms from the security point of view .The issues of information security like confidentiality,
integrity,authenticity etc are taken care of in the field of application in cryptography. Cryptography itself is divided into two parts namely, quantum cryptography and classical cryptography. Quantum cryptography is out of the scope of this paper.
Cryptographic application areas are electronic communication channels, smart cards, electronic currency and defense communications. 
For accomplishing its goal in the information security aspects ,the science of cryptographic techniques are applied to various ciphers which are in existence from many years.Ciphers are those algorithms that are designed by the 
scientists of this field of research.Different ciphers follow different rules of design and structure .There are two kinds of ciphers that exist. Stream as well as Block ciphers. The former allow for encrypting each bit at a time in a stream of input bits and
the later demand encrypting of a block of data rather than just bits.In the next section we will discuss about the subject of block ciphers.

\subsection{Block ciphers}
Block ciphers are the algorithms providing security to certain information through applying specific encryption techniques on a group of bits of the input. These group of  bits are called blocks.
Encryption techniques include the mixing of these input group of bits with the key.This key also consists of a group of bits.Block ciphers are used widely for encryption of large blocks of data.
The first block cipher was introduced by Horst Feistel in 1970s which was named as Lucifer.Block cipher designs include the input to be of a certain bit length and also the length of the key bits is important.
For example, Lucifer later renamed as the popular block cipher DES, consists of 64 bits in the input block and key length of 56 bits. 
Throwing light on the various families of block ciphers, we may discuss about the prominent two  families,that are, Feistel Network and Substitution Permutation Network.
Lucifer or DES follows Feistel structure. Feistel structure involves the division of the input plaintext into two similar length parts .There is one function used in this structure that mixes the key bits with the 
plaintext bits.The right half is combined with the key using the functon and then the output of the function is XORed with the left half.After this, the XORed output of left half becomes right half and the right half becomes the new left half 
without any change. The Substitution Permutation Network  has a two layered structure.First layer 
consists of confusion and the second layer provides diffusion. Confusion and diffusion are the two properties introduced by Claude Shannon in the coding theory, which are used in cryptography. 
Confusion is introduced in the form of substitution and diffusion is introduced in the form of permutation.
Here, in ciphers the confusion is incorporated in the algorithm by usage of S-box . The diffusion part is fulfilled by using some permutaion methods.
Confusion , is the idea of severely changing the data on the way from input to output.Diffusion is the idea of the huge change in the output based on a single change in the input. 
Modern block ciphers follow an iterative structure. This helps in increasing the security of these ciphers.These iterations are in language of cryptography,known as rounds.
Rounds provide security to an algorithm after certain number of its occurrences. It depends on the strength of the  components of confusion and diffusion 
of an algorithm that how many rounds will it require to acquire enough strength.

\begin{figure}[H]

\includegraphics[width=5cm,height=10cm,keepaspectratio]{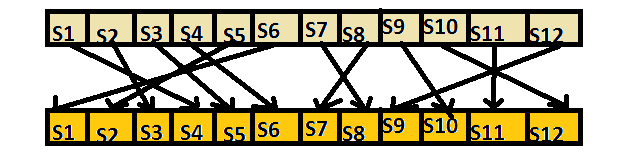}
\centering
\caption{This is the idea of the SPN structure}

\end{figure}

\begin{figure}[H]
\includegraphics[scale=0.3]{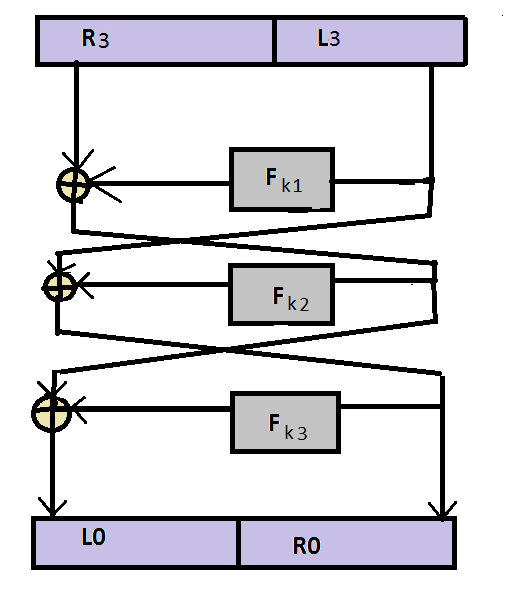} 
\centering
\caption{The diagram above shows the structure of Feistel Network}

\end{figure}

\subsection{Lightweight Block Ciphers}

Lightweight cryptography is the subject of designing lightweight ciphers .Lightweight ciphers are suitable for their performance in more constraint environments.
Its central focus is the better performance of these ciphers with low power consumption.Used for 
 RFID tags, medical and health care devices and devices that use sensors.Lightweight block ciphers is a current research topic for many experts in the field of cryptography.
These lightweight block ciphers are originated from three families of ciphers, namely, Substitution Permutation Network, Feistel Network and Generalized Feistel Network.For a survey on the various ciphers under these categories,we may refer to the paper[10]. 
If we focus on the ciphers belonging to the SPN family, we will find many names such as, AES, LED, PRESENT,PRIDE and more.
Looking at the number of ciphers under the Feistel Network,we can name a few of them, for example, HIGHT,LEA,RoadRunner etc.
While,CLEFIA,TWINE,Piccolo comes under Generalized Feistel Network ciphers.These lightweight block ciphers follow some certain specifications and give high performances in both hardware as well as softwares.
In our paper, we have followed the Substitution Permutation Network for designing the new cipher.There is a branch of ciphers called as the LS designs under the SPN family.
Its two instances are the ciphers Robin and Fantomas[4,6,7]. Both of these are 128 bit ciphers.Our work is based on the extended version of these cipher designs. 
For instance, if we talk about the most recently proposed lightweight block cipher algorithm, we will have to move from LS to XLS i.e Extended LS family of designs.
The lightweight block cipher Mysterion, belongs to the Substitution Permutation Network family of lightweight block ciphers.
It follows a structure which is quite similar to the block cipher AES.This block cipher belongs to the Extended LS design ciphers category which is the extended version of the lightweight block cipher design, that is, namely, LS design.

\section{Our Proposal for the new cipher}\label{sec:Others}
In the recent past, there have been much research on the simplicity of block ciphers and their resistance to more and more attacks.These research has lead to the construction and testing of various lightweight instances.
Our work is focused on the design and construction of such a structure.The main idea is taken from the Extended LS family of designs. Mysterion is the only instance of this family and is available in 128 bit as well as 256 bit block lengths.
The new cipher structure is a step towards designing a 256 bit Extended LS cipher.If we recall the design of LS ciphers Robin and Fantomas[4,6,7], they perform the encryption job on a $8\times16$  block of data.
Robin is the involutive instance and Fantomas[4,6,7] is the non involutive one.They use S-box for the columns i.e, 16 columns and L-box for the rows,i.e, 8 rows. The binary matrices are used as lookup tables in both the instances  
for the L box operation.There are block ciphers which use the binary matrices for their diffusion layers as in the case of ARIA[9].8 bit S boxes and 16 bit L boxes are used from the  involutive and non involutive category for the respective ciphers.LS-Designs were famous for their efficient masked implementation aspects.These kind of implementation 
provides resistance against Side Channel attacks.Bitslice non linear S boxes are combined with linear L boxes in such designs.
Moving towards the XLS family, if we describe the structure and design of Mysterion [2], we will see that its input state is subdivided into four blocks of $4\times8$ dimension.Here also,the S box is applied to the columns and L box[3,8] is applied to the rows all at once  rather than being applied to rows one by one. The Mysterion instance adds an additional ShiftColumns operation after the L box operation ,that distributes the columns of each block to the other blocks.In this way, the diffusion is spread . After the ShiftColumns step, the key and the round constants are added to the state .Mysterion 128, involves the use of 4 subdivisions of the state into blocks and Mysterion 256 involves 8 subdivisions of the state into blocks. 

The algorithm of LS Design is given below:



\begin{algorithm}
\caption{LS Design algorithm with L bit L boxes and s bit S boxes}\label{euclid}
\begin{algorithmic}[1]
\State $X\gets P\oplus K$\Comment{X is an s.l bit matrix}
\For{\texttt{R=0 TO N}}
\For{\texttt{I=0  TO L}}   \Comment{The S box layer}
\State $\texttt{X[I,$\star$]} \gets  \texttt{S[X[$\star$,I]]}$
\EndFor
\For{\texttt{J=0  TO S}}   \Comment{The L box layer}
\State $\texttt{X[$\star$,J]} \gets  \texttt{L[X[J,$\star$]]}$
\EndFor
\State $X \gets \texttt{ $X \oplus K \oplus C(R)$}$  \Comment{Key addition and round constant}
\EndFor

\State \textbf{return} $X$

\end{algorithmic}
\end{algorithm}

In the above algorithm, the definition of  L bit L boxes and s bit S boxes produces an m=s.L bit cipher.

\subsection{The new instance,Marvin}
Our newly proposed cipher follows the LS Design criteria for the columns and rows of the block,whereas,its state is subdivided into 4 blocks
of $4\times16$ dimension.The S box is applied to the columns of each block as in any LS design instance and the L box is applied to the rows of every block as in the
LS designs.The key of this instance is also of the same dimensions as the input state and is XORed with the state at every round. Along with the key, a round constant is also XORed with the state at every round.Additionally,
 an intermediate stage of permutation layer has been constructed that is included in this cipher. It is called as the PermuteSets operation which illustrates the permutation of a part of the columns among the different blocks of the state.
The diagram of the design of the new block cipher Marvin is as given below:

\begin{figure}[H]

\includegraphics[width=14cm,height=15cm,keepaspectratio]{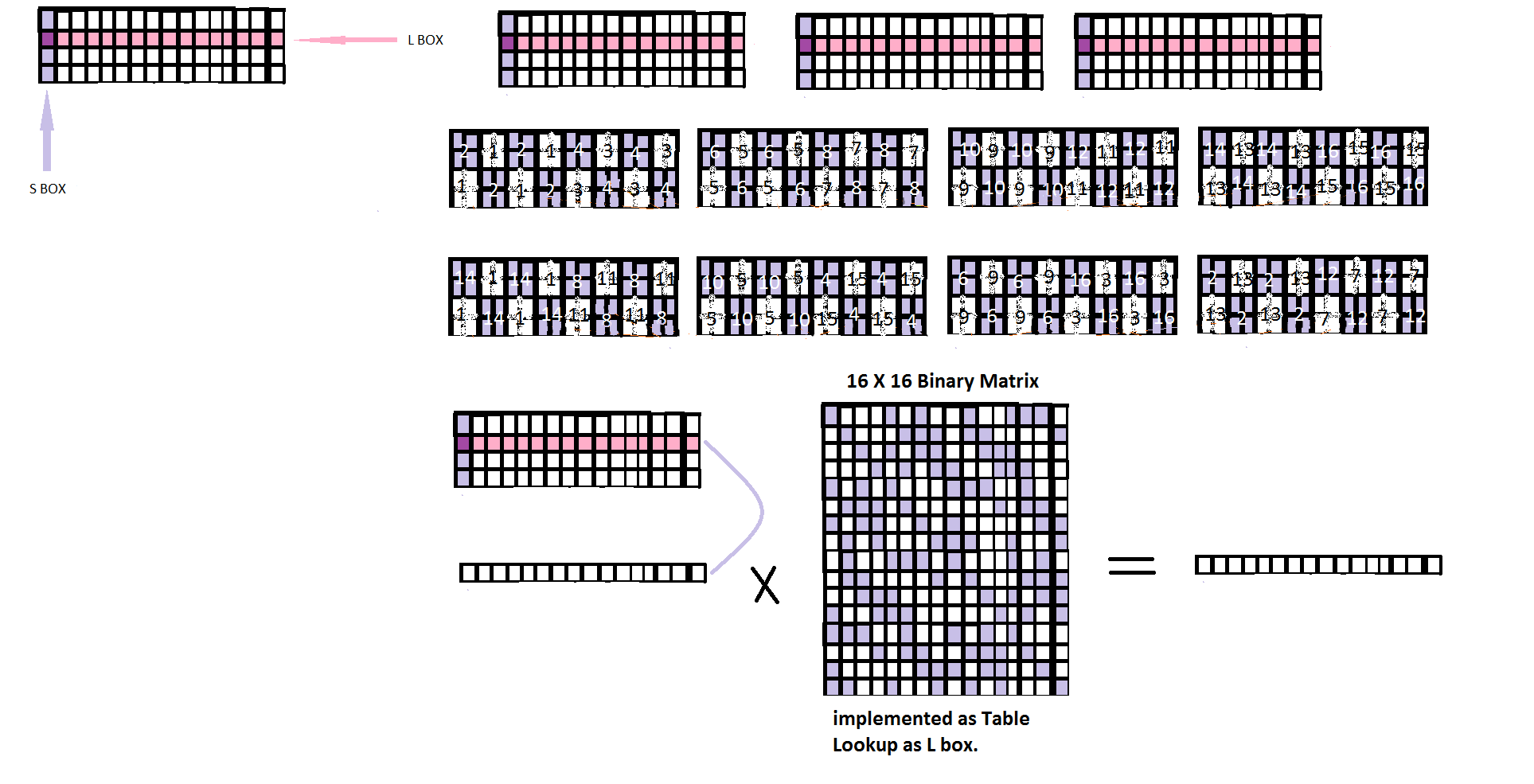}
\centering
\caption{Marvin}
\end{figure}
\subsection{Specifications for Marvin}

\begin{algorithm}
\caption{New XLS Design with b(s.l) bit blocks}
\begin{algorithmic}[1]
\State $X\gets P\oplus K$\Comment{X is a b( s.l ) bit matrix}
\For{\texttt{R=0 TO N}}
\For{\texttt{J=0  TO b}}
\For{\texttt{I=0  TO l}}           \Comment{The S box layer}
\State $\texttt{X[J,$\star$,I]} \gets  \texttt{S[X[J,$\star$,I]]}$
\EndFor
\For{\texttt{K=0  TO s}}           \Comment{PermuteSets  layer}
\State $\texttt{X[$\star$,K,$\star$]} \gets  \texttt{PermuteSets[X[$\star$,K,$\star$]]}$
\EndFor
\For{\texttt{J=0  TO b}}   \Comment{The L box layer}
\State $\texttt{X[J,$\star$,$\star$]} \gets  \texttt{L[X[J,$\star$,$\star$]]}$
\EndFor
\State $X \gets \texttt{ $X \oplus K \oplus C(R)$}$  \Comment{Key and round constant addition}
\EndFor
\EndFor
\State \textbf{return} $X$

\end{algorithmic}
\end{algorithm}

The internal state in the above algorithm can be seen as the X[$\star$,$\star$,$\star$] matrix.
Here, x[i,$\star$,$\star$] is the ith $0 \leq i \leq b$  block of the state ,x[i,j,$\star$] is the jth row of the ith block in which L box has to be applied
and x[i,$\star$,j] is the jth column of the ith block in which the S box has to be applied.The PermuteSets operation works on the x[$\star$,$\star$,k] state,
which permutes the sets from the two consecutive columns of the blocks among the blocks.
Now, we will describe the different components of this cipher Marvin.
It is designed in a way that suits a 256 bit block size and a 256 bit key as well.
\subsubsection{S box}

Marvin uses the 4 bit involutive S box with optimal bitslice representation,that was discovered by the recently published paper by V Grosso, G Leurent, FX Standaert, K Varici [6].
It consists of 4 AND gates and 4 XOR gates.Its algebraic degree is 3 ,with differential probability $2^{-2}$
and linear probability $2^{-1}$.
\begin{algorithm}
\caption{S box Logic operations \textit{ (a,b,c,d are the four input bits)}}
\begin{algorithmic}

\State   $z = b   \&   c $ \\
\State   $x = a \oplus z $\\
\State   $ q = x   \&  c $\\
\State   $ s = q \oplus d$\\
\State   $ m = s   \&   b  $\\
\State   $ t  = m   \oplus  c $\\
\State   $ n = t    \&    s$\\
\State   $ p = n    \oplus  b$\\
\State   $ => a'  =   p$\\
\State   $  => d' = t$\\
\State   $  =>  c' = s$\\
\State   $  => b' = x$\\
\end{algorithmic}
\end{algorithm}

The S box is as follows:
\begin{figure}[H]

\includegraphics[width=8cm,height=9cm,keepaspectratio]{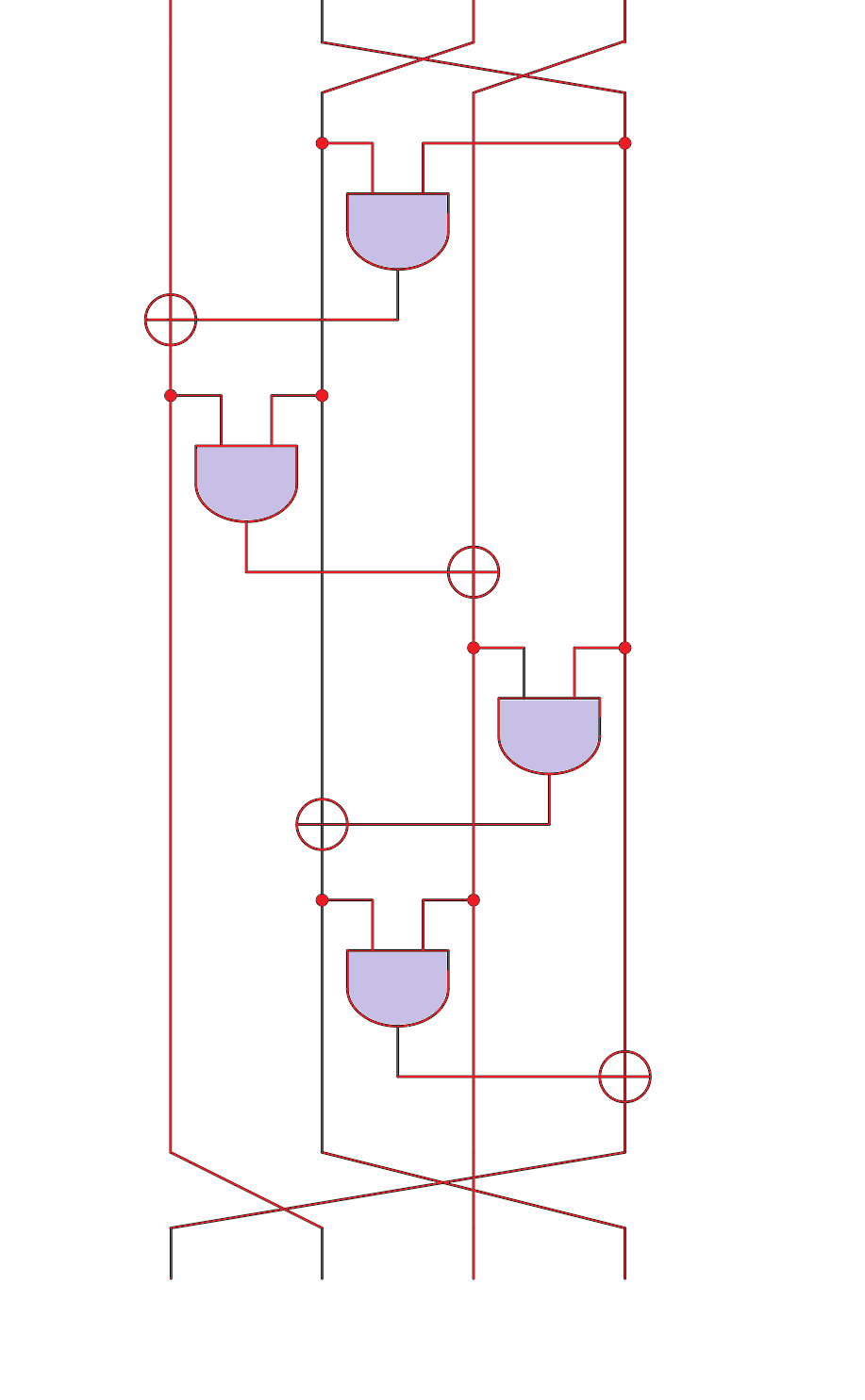}
\centering
\caption{4 bit S box}
\end{figure}      

\subsubsection{PermuteSets }

In the PermuteSets operation, the consecutive pairs of columns are partitioned into two halves of four bits each.
These are named as Sets. In the figure below, we may see that the Set (coloured) consists of such four bit partition from a pair of columns.
We have many  coloured Sets  in the figure. And, we perform the permutation of these Sets .
The upper figure of blocks represent the Sets before permutation, the lower figure of blocks 
represent the Sets after permutation.
In the figure below, the upper state shows coloured Sets of 4 Set having same number. These four Sets from a block 
always move together to the other blocks.And this spreading of Set groups is done across all the four blocks.
The spreading feature is illustrated using the various numbers.These coloured sets are shown to be differently placed in the lower state of the figure.This is the way we 
have tried to illustrate the permutation. 
We may see in the below permutation ,that a change in one bit of any block would affect maximum 4 blocks.
Hence ,the branch number of this diffusion or permutation layer, can be calculated to give the answer as 5.
Marvin cipher uses the following permutation within the 4 blocks of the cipher state:
\begin{figure}[H]

\includegraphics[width=12cm,height=15cm,keepaspectratio]{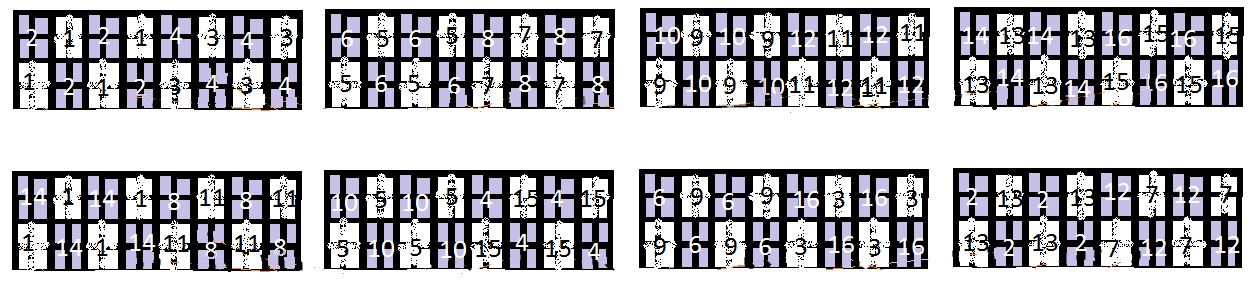}
\centering
\caption{PermuteSets}
\end{figure}      

\subsubsection{L box}
The new cipher Marvin, uses the binary matrix for its diffusion layer. 
Unlike Mysterion,which uses the recursive MDS matrix[3] of branch number 9, we have used the binary matrix derived in the paper by ,B Aslan, MT Sakallı [1].
Ours is a $16\times16$  binary involutive matrix that will be used as a lookup table ,as it is a bitslice cipher.

\begin{figure}[H]
\includegraphics[width=7cm,height=11cm,keepaspectratio]{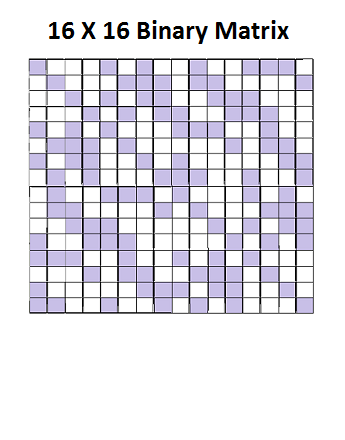}
\centering
The coloured cells represent active (value 1) and the white cells represent inactive (value 0).
\caption{The linear layer}
\end{figure}      
The branch number of this matrix is 8 and it provides better diffusion,if compared to many diffusion layers out there.
Hence, the linear layer we have used is resistant against linear and differential cryptanalysis,as this
provides the optimal branch number.

\section{Analysing the security aspects}

In case of Marvin,as we have designed it for 256 bit block length and the number of block matrix doesnot 
match the number of columns in each block matrix, therefore, the state is not forming a square.
Therefore ,we will not be able to apply the four round propagation theorem of AES directly to lower bound the number 
of active S boxes.
Using the Wide Trail Strategy[5], we can prove the number of rounds for which the cipher will give better security.
We are going to see that how Marvin may satisfy the security bounds with an arrangement in its sequence of round 
transformation.Let us introduce some definitions regarding this analysis.
A bundle is the four bit column in the state of Marvin.We will denote the L box application by L.The partition of 
bundles is called $\Xi$. Marvin is a key alternating block cipher.It applies the same round functions in every iteration.
The round operations are a composition of the application of S boxes to the columns of the state,application of L boxes
to the rows of the state,the application of PermuteSets to the set of bits of the state and addition of key and round constants to the state.
The later donot influence the number of active S boxes.Therefore, in our analysis, we will discard them.
Based on the observation of the round transformations,four rounds of Marvin can be written as,
\[ L \circ PermuteSets \circ S  \circ   L \circ PermuteSets \circ S  \circ   L \circ PermuteSets \circ S  \circ   L \circ PermuteSets \circ S     \]    

Now, we will organise this above sequence of operations in a specific way that will help in the analysis.
We may see that the linear diffusion layer,that is, the L box operation is commutative to the S box operation.
Hence , we may write the above round transformation sequence as,

\[ L  \circ  PermuteSets  \circ  L  \circ  S   \circ  PermuteSets   \circ  S   \circ  L  \circ  PermuteSets   \circ   L  \circ  S   \circ  PermuteSets  \circ  S  \]

Now, from the above sequence, we may derive two different transformations $p^a = PermuteSets  \circ  S  =  \Theta$ 
and  $p^b =  \theta  \circ  S $  where $ \theta = L  \circ  PermuteSets  \circ  L $.  Therefore,four rounds of Marvin is the alternation of $p^a$  and  $p^b$.

\[p^a  \circ  p^b  \circ  p^a  \circ  p^b \]

We have the following theorem from [5].

\textit{ Theorem :  For a key alternating block cipher with round transformation $p^a$  and $p^b$ , the number of active S boxes  of  any trail over
    \[  p^a \circ p^b \circ p^a \circ p^b\]     
is lower bounded by  B($\theta$)  $\times$  B($\Theta,\Xi$) where  B($\theta$)  is the branch number of the linear transformation L and B($\Theta,\Xi$)  is the 
branch number of the linear transformation $\Theta$  with  respect to the partition of the bundles $\Xi$. }

We may calculate the branch number of the binary matrix to know the value of B($\theta$). In our case,it has the branch number 8. The definition of an active block, is an active bundle.
The PermuteSets operation spreads partition Sets of the pair of bundles to other blocks.Hence, one active Set may spread to the four blocks.Therefore,$\Theta$ has a branch number of 5.
Therefore, the number of active S boxes over four rounds is lower bounded by $8 \times 5 = 40$.

From the above proof, we may see that 4 rounds of Marvin will activate at least 40 S boxes.

\[ Pr_{lin}^{char} ( 4 rounds) \leq  Pr_{lin}^{max} ( S box)^{40}  =  2^{-40} \]

\[ Pr_{diff}^{char} ( 4 rounds) \leq  Pr_{diff}^{max} ( S box)^{40}  =  2^{-80} \]

If we set the number of rounds to 28, the probability of differential characteristic will be $2^{-560}$.
Therefore, for Marvin 256, it will guide to comfortable security margins.

\section{Implementation aspects}

Marvin can be implementated in the ARM Assembly Language in Cortex M3 .
The S box operations can be implemented using the logical operators like XOR and AND .
L box layer must be implemented as a lookup table based operation.
The round constants are calculated as in AES algorithm
i.e, in GF($2^8$). The PermuteSets operation can be implemented by logical operations too.
The cipher would take more than 5200 cycles ,if implemented in the ARM assembly.
The cipher designed by us is an involutive one, as it has the involutive S box as well as an involutive binary matrix used 
as the L box. Therefore, Decryption would use the same circuit and the PermuteSets will be just reversed.And the round constants will also be 
applied in the reverse order.

\section{Conclusion}\label{sec:Conclusion}

This work emphasizes the extension of the LS Designs.It gives a new instance Marvin to the 
family of Extended LS designs.Marvin can be used to encrypt a 256 bit block of data, which may take 28 rounds.
 The implementation aspects are more to be explored for better 
search of performance. Moreover, the cipher needs to be tested for more cryptanalytic attacks.


\begin{thebibliography}{10}

\bibitem{Einstein}
Aslan, B. and Sakallı, M.T., 2014.\textit{ Algebraic construction of cryptographically good binary linear transformations.} Security and Communication Networks, 7(1), pp.53-63.\texttt{DOI: 10.1002/sec.556}
\bibitem{Einstein}
Journault, A., Standaert, F.X. and Varici, K., 2017.\textit{ Improving the security and efficiency of block ciphers based on LS-designs.} Designs, Codes and Cryptography, 82(1-2), pp.495-509.\texttt{DOI:10.1007/s10623-016-0193-8}
\bibitem{Einstein}
Augot, D. and Finiasz, M., 2014, March.\textit{ Direct construction of recursive MDS diffusion layers using shortened BCH codes.} In International Workshop on Fast Software Encryption,8540, (pp. 3-17). Springer, Berlin, Heidelberg.\texttt{DOI: 10.1007/978-3-662-46706-0\_1}
\bibitem{Einstein}
Leander, G., Minaud, B. and Rønjom, S., 2015.\textit{ A Generic Approach to Invariant Subspace Attacks: Cryptanalysis of Robin, iSCREAM and Zorro.} EUROCRYPT (1), 9056, pp.254-283.\texttt{DOI:10.1007/978-3-662-46800-5\_11}
\bibitem{Einstein}
Daemen, J. and Rijmen, V., 2001, December.\textit{ The wide trail design strategy.} In IMA International Conference on Cryptography and Coding,2260, (pp. 222-238). Springer, Berlin, Heidelberg.\texttt{DOI:10.1007/3-540-45325-3\_20}
\bibitem{Einstein}
Grosso, V., Leurent, G., Standaert, F.X. and Varıcı, K., 2014, March. \textit{LS-designs: Bitslice encryption for efficient masked software implementations.} In International Workshop on Fast Software Encryption,8540, (pp. 18-37). Springer, Berlin, Heidelberg.
\texttt{DOI: 10.1007/978-3-662-46706-0\_2}
\bibitem{Einstein}
J Cruz, Rafael \& B Reis, Tiago \& Aranha, Diego \& López, Julio \& Kupwade Patil, Harsh. (2017).\textit{ Lightweight Cryptography on ARM}. .\texttt{DOI: 10.13140/RG.2.2.20527.94881.} 
\bibitem{Einstein}
Sajadieh, M., Dakhilalian, M., Mala, H. and Sepehrdad, P., 2012. \textit{Recursive diffusion layers for block ciphers and hash functions.} In Fast Software Encryption,7549, (pp. 385-401). Springer, Berlin, Heidelberg.\texttt{DOI:10.1007/978-3-642-34047-5\_22}
\bibitem{Einstein}
Kwon, D., Kim, J., Park, S., Sung, S.H., Sohn, Y., Song, J.H., Yeom, Y., Yoon, E.J., Lee, S., Lee, J. and Chee, S., 2003, November. \textit{New block cipher: ARIA.} In International Conference on Information Security and Cryptology,2971, (pp. 432-445). Springer, Berlin, Heidelberg.
\texttt{DOI:10.1007/978-3-540-24691-6\_32}
\bibitem{Einstein}
Hatzivasilis, G., Fysarakis, K., Papaefstathiou, I. and Manifavas, C., 2017.\textit{ A review of lightweight block ciphers.} Journal of Cryptographic Engineering, pp.1-44.\texttt{DOI:10.1007/s13389-017-0160-y}

\end{thebibliography}
\end{document}